\documentclass[9pt]{article}  \usepackage{times}
\usepackage{graphicx}

\usepackage{color}

\usepackage[round,numbers,sort&compress]{natbib} 

\usepackage{amsmath}
\usepackage{amssymb}
\usepackage{stackrel}
\usepackage{chemarrow}
\usepackage{graphicx}
\usepackage{textgreek}

\usepackage{booktabs}
\usepackage{dcolumn}
\usepackage{rotating}
\usepackage{multirow}

\renewcommand\appendix{\par
  \setcounter{section}{0}
  \setcounter{subsection}{0}
  \setcounter{figure}{0}
  \setcounter{table}{0}
  \renewcommand\thesection{Appendix \Alph{section}}
  \renewcommand\thefigure{\Alph{section}\arabic{figure}}
  \renewcommand\thetable{\Alph{section}\arabic{table}}
}

\begin{document}

\twocolumn[{\LARGE \textbf{Melting transitions in biomembranes\\*[0.2cm]}}
{\large Tea Mu\v{z}i\'{c}, Fatma Tounsi, S\o{}ren B. Madsen, Denis Pollakowski, Manfred Konrad and Thomas Heimburg\\*[0.1cm]
{\small Niels Bohr Institute, University of Copenhagen, Blegdamsvej 17, 2100 Copenhagen \O, Denmark}\\*[-0.1cm]

{\normalsize \textbf{ABSTRACT}\hspace{0.5cm} We investigated melting transitions in biological membranes in their native state that include their membrane proteins. These membranes originated from \textit{E. coli}, \textit{B. subtilis}, lung surfactant and nerve tissue from the spinal cord of several mammals. For some preparations, we studied the pressure, pH and ionic strength dependence of the transition. For porcine spine, we compared the transition of the native membrane to that of the extracted lipids. All preparations displayed melting transitions of 10-20 degrees below physiological or growth temperature, independent of the organism of origin and the respective cell type. The position of transitions in \textit{E. coli} membranes depends on the growth temperature. We discuss these findings in the context of the thermodynamic theory of membrane fluctuations that leads to largely altered elastic constants, an increase in fluctuation lifetime and in membrane permeability associated with the transitions. We also discuss how to distinguish lipid transitions from protein unfolding transitions. Since the feature of a transition slightly below physiological temperature is conserved even when growth conditions change, we conclude that the transitions are likely to be of major biological importance for the survival and the function of the cell. 
\\*[0.3cm] }}
\noindent\footnotesize{\textbf{Keywords:} thermodynamics; fluctuations; E.coli; lung surfactant; B.subtilis; nerves; elastic constants\\*[0.1cm]}
\noindent\footnotesize {$^{\ast}$corresponding author, theimbu@nbi.ku.dk. }\\
\vspace{0.3cm}
]

\normalsize

\section*{Introduction}\label{introduction}
Lipid bilayers display melting transitions at a temperature T$_m$, during which both lateral and chain order change accompanied by the absorption of heat. The typical range of transition temperatures is from -20$^\circ$ to +60$^\circ$C depending on chain length, chain saturation and the chemical nature of the head groups \cite{Heimburg2007a}. Melting transitions can easily be observed with calorimeters and various spectroscopic methods such as infrared spectroscopy or magnetic resonance. Mixtures of lipids with different melting temperatures display phase behavior that can be plotted in phase diagrams \cite{Lee1977}. These diagrams show the coexistence of phases as a function of both molar fractions of the components and intensive variables such as temperature and pressure. From the theoretical analysis of phase diagrams one can obtain melting profiles of the lipid mixtures, and understand the phase separation processes \cite{Heimburg2007a, Graesboll2014}. Consequently, one finds domain formation in certain re\-gimes of the phase diagram \cite{Korlach1999, Bagatolli1999}. 

It is little known that biological membranes also display melting transitions close to the physiological temperature re\-gime. Several publications in the 1970s reported, for example, melting phenomena in the membranes of \textit{Mycoplasma laidlawii} \cite{Steim1969, Reinert1970}, \textit{Micrococcus lysodeikticus} \cite{Ashe1971}, mouse fibroblast LM cells \cite{Wisnieski1974}, and red blood cells \cite{Chow1981}. Haest et al. \cite{Haest1974} showed by electron microscopy that the arrangement of proteins in a native bacterial membrane correlates with this transition. Transitions have also been reported for lung surfactant \cite{Nag1998} and \textit{E.coli} membranes \cite{Nakayama1980}. However, the relevance of these transitions for the function of cells has been little appreciated. Since biological membranes consist of hundreds or even thousands of components \cite{White1973, Jamieson1977}, it seems impossible to construct phase diagrams. Due to Gibbs phase rule, the possible number of coexisting phases is of similar order as the number of components when neglecting the phase boundaries. The compositions of domains cannot easily be derived from simple physical considerations or measurements if finite size domains are considered. We show below that it is nevertheless possible to extract useful information from the melting profiles.

Melting transitions strongly influence elastic constants \linebreak\cite{Heimburg1998, Ebel2001}. According to the fluctuation-dissipation theorem, the heat capacity is  proportional to enthalpy fluctuations. All other susceptibilities are similarly related to fluctuations of extensive variables. For instance, the isothermal area compressibility is proportional to area fluctuations \cite{Heimburg1998, Heimburg2005c} and the capacitance is proportional to charge fluctuations \cite{Heimburg2012}. Further, the isothermal compressibility is proportional to the heat capacity \cite{Heimburg1998, Ebel2001, Schrader2002}. The same is true for the other susceptibilities. They are all related to the heat capacity. Consequently, both heat capacity, compressibility and capacitive susceptibility are at a maximum in the melting transition \cite{Heimburg1998, Heimburg2012}. Heat capacity and area compressibility are also related to the bending elasticity \cite{Evans1982, Heimburg1998, Dimova2000}, i.e., they are also at a maximum in a transition. The bending elasticity, however, is an important property for the fusion and fission of membranes \cite{Kozlovsky2002}. This implies that endocytotic and exocytotic events are potentially enhanced in transitions.

Furthermore, the area compressibility is related to the permeability of membranes \cite{Papahadjopoulos1973,  Sabra1996, Blicher2009, Heimburg2010}. To create pores, the membrane in their surroundings has to be locally compressed, \linebreak which is facilitated close to transitions \cite{Nagle1978b, Heimburg2010}. The pores in the membrane are also related to the formation of lipid ion channels, i.e., pores in the lipid membranes that display conduction patterns that are practically indistinguishable from protein channels \cite{Blicher2009, Wodzinska2009, Wunderlich2009, Heimburg2010, Laub2012, Blicher2013, Mosgaard2013b}. The fluctuation-dissipation theorem implies that large fluctuations go along with longer fluctuation lifetimes. This results in longer mean open-lifeti\-mes of lipid pores in the transition range \cite{Grabitz2002, Seeger2007}. The lifetimes of lipid membrane-fluctuations span the range from milliseconds to seconds, and are therefore just in the range observed for protein channel open-lifetimes.

It seems likely that also dynamics properties in membra\-nes are related to transitions. For instance, the sound velocity in lipid dispersions is a function of the membrane compressibility. As a consequence, sound velocities in transitions are reduced \cite{Halstenberg1998, Schrader2002}. Based on this observation, it has been proposed that the presence of a phase transition gives rise to the possibility of the propagation of solitary pulses (solitons) in cylindrical membranes that resemble action potentials \cite{Heimburg2005c, Heimburg2007b, Andersen2009, Lautrup2011}. 

Various authors have shown that transitions are influenced by drugs, e.g., by anesthetics, neurotransmitters or peptides \cite{Seeger2007}. Anesthetics lower the transition temperature of lipids by a well-known mechanism called melting-point depression \cite{Kaminoh1992, Kharakoz2001, Heimburg2007c, Graesboll2014}. The observed pressure-reversal of anesthesia is well explained by the influence of hydrostatic pressure on melting transitions \cite{Trudell1975, Kamaya1979, Heimburg2007c}. Within the soliton theory for the nerve pulse, the effect of anesthetics is explained by the increased free energy threshold for the induction of a phase transition \cite{Wang2018}.

The striking influence of the lipid transition on all kinds of membrane properties which are of biological relevance suggests a careful evaluation of this phenomenon. In this work we investigate the melting in various biological membranes, including the bacteria \textit{Escherichia coli} and \textit{Bacillus subtilis}, lung surfactant, and nerve preparations from rat, sheep and pig. We show that the transitions in these systems are all very similar and are found 10-20 degrees below growth or body temperature. If the growth temperature of bacteria is changed, the transitions shift as well in the same direction. We investigate the role of pressure and discuss the lifetimes of membrane perturbations. In the Discussion, we address the question of the putative role of such transitions in biology.

\section*{Materials and Methods}\label{Methods}
\textbf{Lipids:} Lipids were purchased from Avanti Polar Lipids \linebreak(Birmingham, AL) and used without further purification.  Bo\-vine lung surfactant (BLES Biochemicals Inc., London, Ontario) was a gift from Prof Fred Possmeyer (London, Western Ontario). BLES (bovine lipid extract surfactant) contains small amounts of membrane soluble proteins (SP-B and SP-C) and it contains 77 weight\% zwitterionic lipids. More than half of it is dipalmitoyl phosphatidylcholine (DPPC, 41\% of total weight). The exact composition of BLES is given in \cite{Zhang2011}. CUROSURF (Chiesi Limited, Manchester, UK) was a gift from S\o{}ren Thor Larsen from Haldor Topsoe A/S (Denmark). It is prepared from porcine lungsurfactant, and contains around 70-76 weight \% zwitterionic phospholipids, \linebreak about two-thirds of which is DPPC. The hydrophobic proteins SP-B and SP-C represent about 1\% of the total weight. Details of the composition are given in \cite{Taeusch2002}.

\textbf{Bacterial cells:} The \textit{E.\,coli} strain XL1 blue with tetracycline resistance (Stratagene, La Jolla, CA) and \textit{Bacillus subtilis} were grown in an LB-medium at 37\,$^{\circ}$C.  Bacterial membranes were  disrupted in a French Press at 1200 bar (Gaulin, APV Homogeniser GmbH, L\"{u}beck, Germany) and centrifu\-ged at low speed in a desk centrifuge to remove solid impurities.  The remaining supernatant was centrifuged at high speed in a Beckmann ultracentrifuge (50000 rpm) in a Ti70 rotor, or in a fast desktop centrifuge, to separate the membranes from soluble proteins and nucleic acids.  The pellet was resuspended in buffer (33 vol\% glycerol plus 67 vol\% 10mM Tris, 1mM EDTA, pH7.2) and centrifuged again. The membrane fractions in the pellets were measured in a calorimeter.  The concentration of membranes in the calorimetric \textit{E. coli} was 26.3 mg/ml as determined from the dry weight of the samples. Lipid melting peaks and protein unfolding profiles can easily be distinguished in pressure calorimetry due to their characteristic pressure dependences, the pressure dependence of lipid transitions being much higher than that of proteins \cite{Ebel2001}. 

\textbf{Rat central brain:} Rat brains were donations from the Rigshospitalet in Copenhagen (Prof. Niels V. Olsen). They were kept in a freezer until use. We used the central brain and parts of the spinal cord. The central brain tissue was ground with mortar and pestle. The resulting liquid sample was diluted in a buffer (150 mM KCl, 3mM Hepes, 3 mM EDTA, pH 7.2-7.4). Subsequently, the sample was  sonicated with a high-power ultrasonic cell disruptor (Branson Ultrasonic cell disruptor B15, Danbury, CT, USA) in pulse mode to prevent heating of the sample. Finally, water soluble parts were washed away from the sample by centrifugation. The pellet was assumed to mainly consist of membranes as was apparent in pressure calorimetry (see results). It was used for the calorimetric experiments. More details can be found in \cite{Madsen2011}

\textbf{Mammalian spines:} Sheep spines were bought from a local butcher and were kept on ice. The spine was opened with a saw and the spinal cord was removed. Using scissors and tweezers the dura mater which surrounds the spinal cord was removed. The spinal cord was cut into small pieces and homogenized over a period of 20 minutes with a stator rotor (Tissue Master 125 W Lab Homogenizer, Omni International, Inc, Kennesaw, GA) at 33.000 RPM with a 7mm probe head in 30-second intervals with breaks of 30 seconds to prevent heating. The homogenate was dissolved in a buffer (150 mM NaCl, 1mM Hepes, 2mM EDTA,  pH 7.4) and spun down in a desk centrifuge at 3360 g for 15 minutes. The sample was centrifuged using an MSE Super Minor centrifuge (England) at 3355 RCF in 15 min intervals. After each round the supernatant was discarded, tubes were filled up to the previous level with buffer, vortexed and put in the centrifuge for another cycle until the supernatant was completely clear. The majority of the fibrous tissue that sediments at the bottom was removed by pouring the viscous pellet into a new tube after each centrifugation round.

Porcine spines were bought from the local butcher, and a homogenate of spinal cord tissue was prepared as for sheep spines. The homogenate was filtrated through a stainless steel 100 mesh with 140 \textmu m opening size (Ted Pella, Inc, Redding, CA) in order to remove fibrous tissue. The homogenate was dissolved in a 150mM NaCl 11.8 mM phosphate buffer, pH 7.4 and treated as in the sheep spine preparation. Details on sheep and porcine spine preparations can be found in \cite{Tounsi2015, Muzic2016}.

\textbf{Calorimetry:} Heat capacity profiles were obtained using a VP scanning-calorimeter
(MicroCal, Northampton, MA) at scan rates of 20\,deg/hr (for spinal cord of rats, sheep, and chicken) or 30\,deg/hr for lung surfactant, \textit{E.coli} and \textit{Bacillus subtilis} membranes. This is much faster than the scan rate we typically use for pure lipids (typically $\leqslant$ 5\,deg/hr). This is justified if the expected melting profiles are very broad, and the c$_p$ maximum values are small. A faster scan rate increases the power of the calorimetric response and thus the strength of the signal. The small magnitude of the heat capacity leads to fast relaxation behavior \cite{Grabitz2002, Seeger2007} which enables us to scan fast without hysteresis problems. Pressure calorimetry was performed in a steel capillary inserted into the calorimetric cell as previously described \cite{Ebel2001, Grabitz2002}. In these experiments, absolute heat capacities are not given. In the porcine spine preparations, 30 vol\% glycerol was added to the sample solution in order to prevent freezing of the sample at temperatures below 0$^\circ$C in the calorimeter. A crucial procedure in the analysis of heat capacity profiles with broad and weak signal is the subtraction of a baseline. It is shown in the supplementary information.

\section*{Theoretical considerations}\label{Theory}

In the following sections we outline why a maximum in heat capacity in a biomembrane melting transition is important for its physical properties. We show how it determines the membrane compressibility, its elasticity and the lifetime of membrane perturbation which are relevant for the spontaneous \linebreak opening of membrane pores.  

\subsection*{Fluctuations, susceptibilities and fluctuation lifetimes}\label{fluctuations}

According to the fluctuation-dissipation theorem, the heat capacity of a membrane is related to enthalpy fluctuations \cite{Heimburg1998} through
\begin{equation}\label{eq:theory01}
	c_p=\left( \frac{\partial H}{\partial T}\right)_p=\frac{\left\langle H^2\right\rangle -\left\langle H\right\rangle ^2}{kT^2}
\end{equation}
Similarly, the isothermal volume compressibility is related to volume fluctuations \cite{Heimburg1998}
\begin{equation}\label{eq:theory02}
	\kappa_T^V =-\frac{1}{\left\langle V\right\rangle } \left( \frac{\partial V}{\partial p}\right)_T=   \frac{\left\langle V^2\right\rangle -\left\langle V\right\rangle ^2}{\left\langle V\right\rangle\cdot kT}
\end{equation}
and the isothermal area compressibility is related to area fluctuations.
\begin{equation}\label{eq:theory03}
	\kappa_T^A =-\frac{1}{\left\langle A\right\rangle } \left( \frac{\partial A}{\partial \Pi}\right)_T=   \frac{\left\langle A^2\right\rangle -\left\langle A\right\rangle ^2}{\left\langle A\right\rangle\cdot kT} \;.
\end{equation}
It is a phenomenological finding that for DPPC and some other lipids \cite{Ebel2001, Seeger2007}
\begin{equation}\label{eq:theory04}
	V(T)\approx \gamma_V H(T)\qquad ; \qquad A(T)\approx \gamma_A H(T)
\end{equation}
where $\gamma_V=7.8\cdot 10^{-10}$m$^2$/N \cite{Ebel2001} and $\gamma_A=0.893$m/N \cite{Heimburg1998} for DPPC. It has been shown that the parameters $\gamma_V$ and $\gamma_A$ are very similar for different lipids, lipid mixtures and even biological preparations such as lung surfactant. Using eqs. (\ref{eq:theory01})-(\ref{eq:theory04}), one can conclude that 
\begin{equation}\label{eq:theory05}
	\kappa_T^V =\frac{\gamma_V^2\cdot T}{\left\langle V\right\rangle }c_p \qquad ; \qquad \kappa_T^A =\frac{\gamma_A^2\cdot T}{\left\langle A\right\rangle }c_p
\end{equation}
i.e., the excess compressibilities are proportional to the excess heat capacity changes. Molecular dynamics simulations suggest that these relations are also true for absolute heat capacities and the total compressibilities \cite{Pedersen2010}.\\

Through the fluctuation-dissipation theorem, the fluctuations are also coupled to the relaxation time $\tau$,
\begin{equation}\label{eq:theory06}
	\tau =\frac{T^2}{L}c_p
\end{equation}
where $L\approx 7\cdot 10^8$ J$\cdot$K /mol$\cdot$s \cite{Grabitz2002, Seeger2007}. This implies that regions of high heat capacity display slow relaxation processes. Relaxation times are identical to fluctuation lifetimes \cite{Onsager1931b}. Therefore, it has been suggested that these lifetimes correspond to the open-lifetimes of lipid ion channels, which are pores in the membrane with a conductance signature that is indistinguishable from that of a protein ion channel \cite{Heimburg2010, Mosgaard2013b}. It has in fact been demonstrated experimentally that the channel lifetimes are related to the heat capacity of the membrane for lipid membranes, but also for protein channels reconstituted into synthetic membranes that display a transition close to experimental temperature \cite{Seeger2010, Mosgaard2013b}. 

\subsection*{Pressure dependence of transitions and volume compressibility}\label{pressure}
The pressure dependence of lipid and protein transitions is very different. Membranes generally increase their transition temperatures upon increase of hydrostatic pressure because the excess volume of the membrane is of the order of +4\% and the melting temperature is roughly given by $T_m=(\Delta E+p\Delta V)/\Delta S$, i.e., it is proportional to the hydrostatic pressure. Protein unfolding transitions, in contrast, usually display a very small excess volume which is negative \cite{Royer2002, Ravindra2003}. A negative excess volume implies that pressure lowers the unfolding temperature of proteins. Therefore, they can be denatured by high pressure. This implies that lipid and protein transitions can be distinguished in calorimetric scans performed at different pressures.

For lipid membrane transitions it has been shown that if eq. (\ref{eq:theory04}) is valid for all temperatures, one can deduce the same information from heat capacity profiles obtained in the presence of a hydrostatic pressure difference, $\Delta p$.  The enthalpy $\left\langle \Delta H\right\rangle _T^{\Delta p}$ obtained at excess pressure $\Delta p$ can be superimposed with an enthalpy profile obtained at $\Delta p=0$ when the temperature $T$ is rescaled to a new temperature $T^\ast$ according to the following relations \cite{Ebel2001}:
\begin{eqnarray}\label{eq:theory07}
	\left\langle \Delta H\right\rangle_T^{\Delta p}&=&(1+\gamma_V\cdot \Delta p)\left\langle \Delta H\right\rangle_{T^\ast}^{\Delta p=0}\\
	&\mbox{with}&T^\ast=\frac{T}{1+\gamma_V\cdot \Delta p}\equiv f\cdot T \;,\nonumber
\end{eqnarray}
This relation allows us to check two properties of lipid melting: 1. a proportional relation between excess volume and enthalpy holding for all temperatures and 2. the value of $\gamma_V$. If eq. (\ref{eq:theory07}) leads to two superimposable $c_p$ profiles, one can conclude that relations (\ref{eq:theory05}) are also valid for biological preparations. We will use this relation below to determine $\gamma_V$ for a series of biological preparations, and to confirm that the proportional relation between excess compressibility and heat capacity also holds for biological melting transitions.
\begin{figure*}[htb!]
	\centering
	\includegraphics[width=14.0cm]{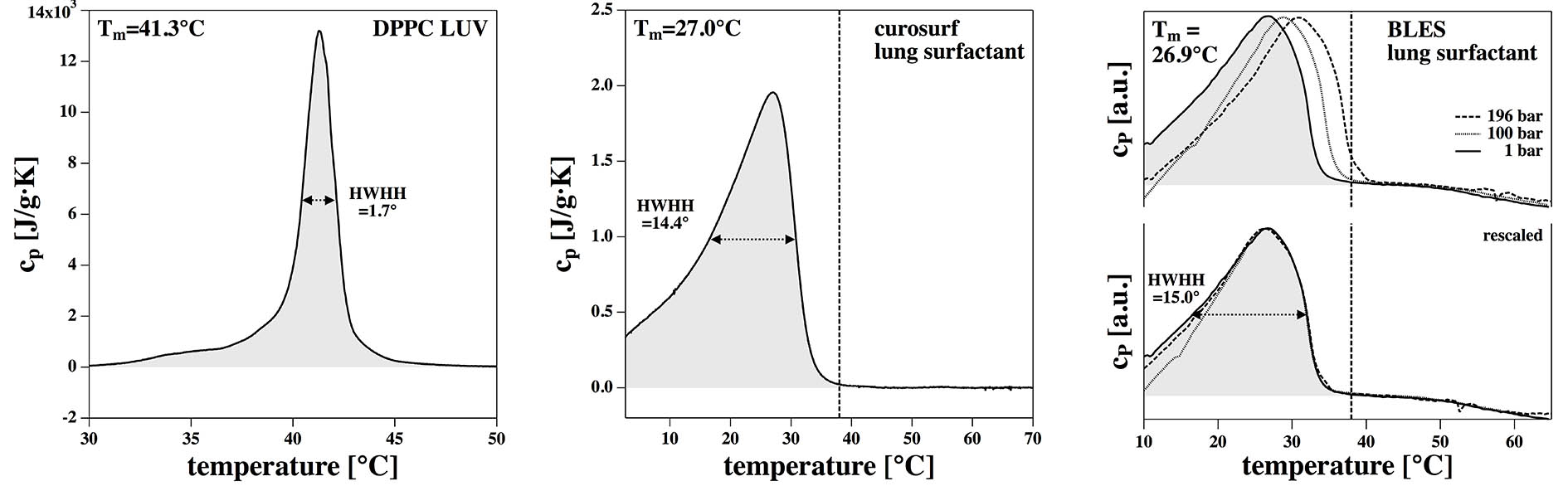}
	\parbox[c]{13cm}{ \caption{\textit{Heat capacity profiles of lung surfactant. Left: DPPC large unilamellar vesicles (LUV). The half width of the peak is 1.7$^\circ$. Center: Curosurf, a lipid extract from porcine lungs. Half width is 14.4$^\circ$.  Right: Bovine lipid extract surfactant (BLES) as a function of pressure (top panel) and rescaled using eq.\ref{eq:theory07} (bottom panel). The half width is 15.0\,$^\circ$. Both lung surfactant preparations show a transition peak with a transition temperature of 26.9\,$^\circ$C. }
			\label{Figure_lung_surfactant}}}
\end{figure*}

A similar relation for the dependence of membranes on lateral pressure is very likely but more difficult to measure. There exists indirect evidence that the proportional relation between enthalpy and area, and the second relation in eq. (\ref{eq:theory05}) are also correct \cite{Evans1982, Heimburg1998, Dimova2000, Pedersen2010}.

\subsubsection*{Reversibility of protein unfolding and lipid melting}\label{denaturation}
Most proteins unfold  irreversibly upon heating, which is a consequence of the aggregation of unfolded chains that expose hydrophobic residues to water. Since aggregation is a slow process, protein unfolding may be partially reversible on a short time scale or in consecutive scans. In contrast, lipid melting is always fully reversible. This allows us to distinguish protein unfolding from lipid melting in several consecutive heating scans in the calorimeter. 

\section*{Experimental Results}\label{Results}
Here, we present studies on various types of cell membranes. These include two different lung surfactant preparations, \textit{E. coli} and \textit{B. subtilis} membranes, and three different brain and spine preparations from rat, sheep and pig.

\subsection*{Lung surfactant}\label{lung}
Lung surfactant is a lipid film that exists in a monolayer-bilayer equilibrium on the surface of the alveoli of the lung \cite{PerezGil2010, Echaide2017}. Its purpose is to reduce the surface tension at the air-water interface and to prevent the lung from collapsing. It contains about 5\% of surfactant-associated proteins (SP-A, B, C and D). The rest is lipids, predominantly DPPC ($\approx$40\%) with a melting temperature of 41.2$^\circ$C. In clinical applications, one often uses lipid extracts from the surfactant. It reduces the surface tension of the air-water interface in the alveoli and prevents the lung from collapsing due to the capillary effect.  Two commercial surfactants are bovine lipid extract surfactant (BLES) prepared from bovine lungs and Curosurf extracted from porcine lungs containing about 2\% hydrophobic proteins \cite{Zhang2011}. It consists mostly of phospholipids with minor contents of the hydrophobic proteins SP-B and SP-C. Since these preparations are close to native membrane preparations, nearly free of proteins,  and available in larger quantities, they are a good starting material for the study of transitions. Due to the high content in DPPC, they can readily be compared to a pure DPPC dispersion.
\begin{figure*}[htb!]
	\centering
	\includegraphics[width=14.0cm]{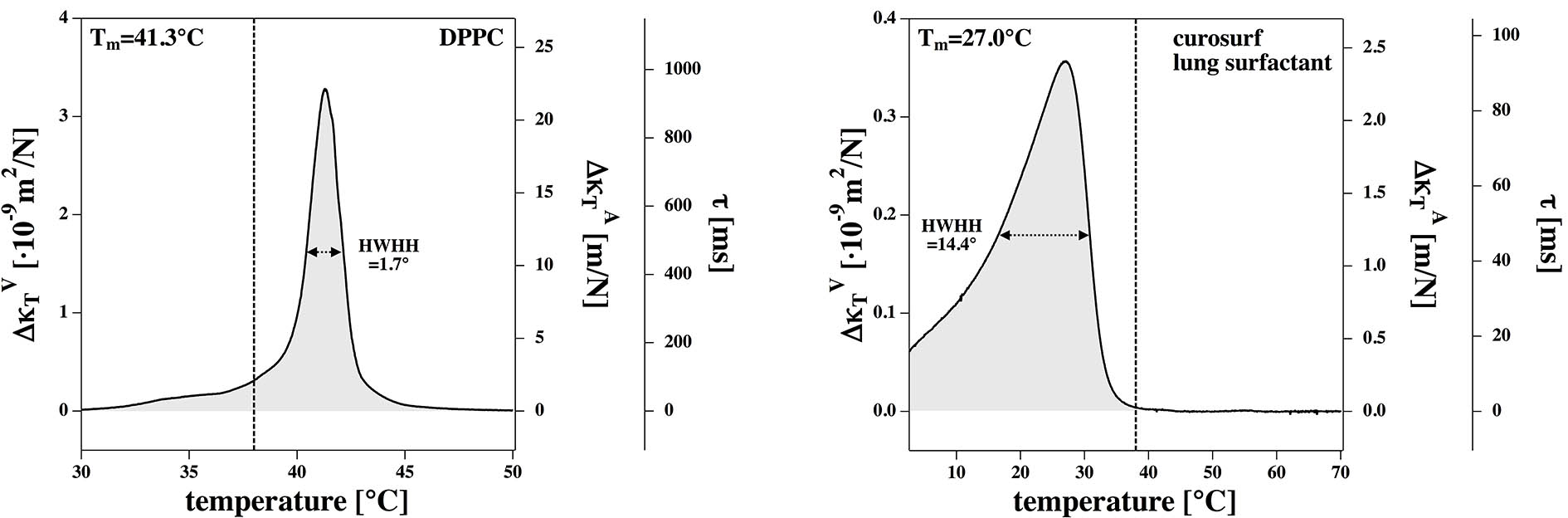}
	\parbox[c]{8cm}{ \caption{\textit{Elastic constants and relaxation time scales calculated for DPPC large unilamellar vesicles (LUVs, left) and Curosurf lung surfactant (right).}
			\label{Figure_elast_DPPC_lung_curosurf}}}
\end{figure*}
Fig. \ref{Figure_lung_surfactant} shows the heat capacity profiles of DPPC large unilamellar vesicles (LUV) (Fig. \ref{Figure_lung_surfactant}, left), Curosurf lung surfactant extract in units of J/g$\cdot$K (Fig. \ref{Figure_lung_surfactant}, center) and BLES (Fig. \ref{Figure_lung_surfactant},right, top panel) in arbitrary units at three different pressures (1 bar, 100 bar and 196 bar). The integral of the $c_p$-profile of Curosurf yields $\Delta H \approx$ 32 J/g of surfactant. The true value is somewhat higher because the $c_p$-profile extends to a temperature below zero which is not accessible in our experiment. For comparison, the major lipid component of lung surfactant (DPPC) possesses a melting enthalpy for the main transition of $\Delta H=$45 J/g, which is of similar order as the surfactant melting enthalpy. Therefore, we will in the following assume that the transition enthalpies of DPPC and of lung surfactant are similar.

The right hand panel of Fig. \ref{Figure_lung_surfactant} shows the pressure dependence of the transition profile of BLES. The transition maximum and the half width of the profile at 1 bar are nearly identical to that of Curosurf. The transition peaks shift towards higher temperatures upon increasing pressure while maintaining the shape of the profile. One can multiply the absolute temperature axis with a factor $f$ in order to make the profiles recorded at the three different pressures overlap (Fig. \ref{Figure_lung_surfactant} right, bottom). The respective factors are $f=0.9925$ for the 100 bar recording, and $f=0.985$ for the 196 bar recording. Using eq. (\ref{eq:theory07}), one can now determine a value for $\gamma_V$,
\begin{equation}\label{lung_1}
	\gamma_V=\frac{1-f}{f\cdot \Delta p} \;.
\end{equation} 
This calculation yields $\gamma_V=7.6\cdot 10^{-10}$ m$^2$/N for the 100 bar measurement and $\gamma_V=7.8\cdot 10^{-10}$ m$^2$/N for the 196 bar measurement. This is within error identical to the value determined for DPPC ($\gamma_V=7.8\cdot 10^{-10}$ m$^2$/N) obtained by Ebel et al. \cite{Ebel2001} (see this reference also for an estimation of the errors). This will allow us to make estimates for volume and area compressibilities of lung surfactant (see below), and the relaxation times following eqs. (\ref{eq:theory05}) and (\ref{eq:theory06}). 

For Curosurf, we obtained a melting enthalpy comparable to that of DPPC LUV, and the factor $\gamma_V$ was found to be nearly identical for DPPC LUV and BLES. Taking into account that additionally the transition maximum and the half width of the two lung surfactant preparations are nearly identical, we can make the following assumptions: The melting enthalpy and the factor $\gamma_V=7.8\cdot 10^{-10}$ m$^2$/J are the same for the two lung surfactant preparations and DPPC. We assume that this is also true for the relation between area changes and enthalpy changes. We therefore estimate that $\gamma_A=0.893$ m/N for the three preparations, and that the phenomenological constant in eq. (\ref{eq:theory06}) is given by $L=7\cdot 10^8$J$\cdot$ K/mol$\cdot$s. We further assume that for specific volume and area, the values for DPPC listed in \cite{Heimburg1998} are reasonably close to the values of the biological preparation.

We can now calculate the excess volume compressibility, $\kappa_T^V$, the area compressibility, $\kappa_T^A$, and the (excess) relaxation times $\tau$ in the transition of Curosurf (where the absolute heat capacity values are known); they are given in Fig. \ref{Figure_elast_DPPC_lung_curosurf} in comparison to the DPPC LUV preparation. As shown above, the compressibilities and the relaxation times are roughly proportional to the excess heat capacity. For the maximum volume compressibility of DPPC LUV we find $\Delta\kappa_T^V=33\cdot 10^{-10}$ m$^2$/N, for the area compressibility $\Delta\kappa_T=22$ m/N and for the relaxation time $0.94$ s, respectively. For Curosurf, we find $\Delta\kappa_T^V=3.6 \cdot 10^{-10}$ m$^2$/N, for the area compressibility $\Delta\kappa_T=2.4$ m/N and for the relaxation time $0.093$ s. This implies that there is roughly a factor of 10 between DPPC LUV and lung surfactant, which is also reflected by the finding that the half width of the $c_p$ profile is about 10-fold larger for Curosurf than it is for DPPC LUV.

At physiological temperature, the lipid system (Curosurf) is above its melting temperature and the excess heat capacity is close to zero. This implies that the relaxation times in the lung surfactant preparation are in the millisecond regime. In the Discussion section we will argue that this time scale is related to the time scale of lipid ion channels and the time scale of the nerve impulse.  


\subsection*{\textit{Escherichia coli} and \textit{Bacillus subtilis} membranes}\label{Ecoli}

In a second step, we prepared \textit{E. coli} and \textit{Bacillus subtilis} membranes as indicated in the Methods section. These \textit{E.coli} cells have a tetracycline resistance, and the cultivation medium contains tetracycline to prevent growth of other species. The pelleted membranes were dissolved in a buffer (see Materials section), filled into a calorimeter and scanned with 30 deg/hr. 
\begin{figure*}[tb!]
	\centering
	\includegraphics[width=14cm]{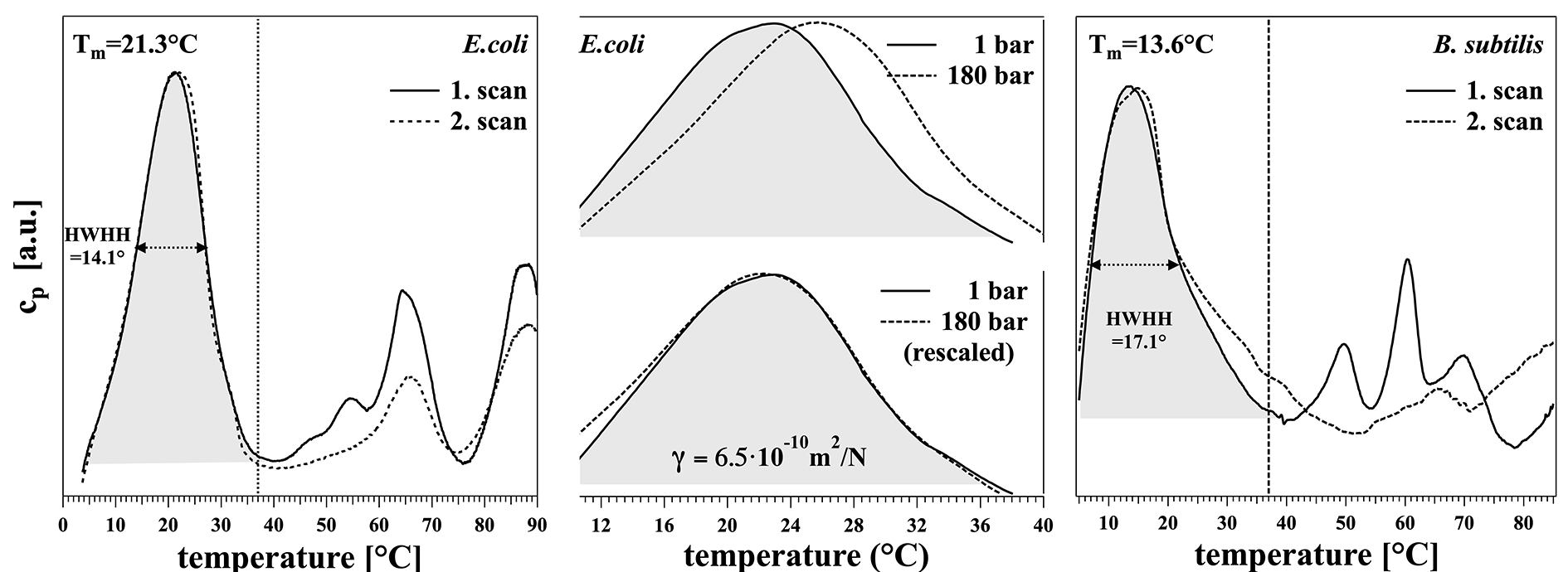}
	\parbox[c]{13cm}{ \caption{\textit{Left: Heat capacity scans of \textit{E.coli} membranes. In the second scan, the protein peaks are largely reduced. The lipid melting peak (grey shaded) is nearly identical on the first and the second scan. Center: The lipid melting peak at two different hydrostatic pressures (1 bar and 180 bar, top panel). The bottom panel shows that the $c_p$-profile rescaled according to eq. (\ref{eq:theory07}) yields two nearly superimposable peaks with $\gamma_V=6.5\cdot 10^{-10}$n$^2$/N, which is very similar to the value for artificial lipids (DPPC: $7.8\cdot 10^{-10}$n$^2$/N).  Right: First and second calorimetric scan of \textit{B. subtilis} membranes. Due to irreversible denaturation, the protein peaks disappear in the second scan. }
			\label{Figure1}}}
\end{figure*}
In a first experiment, \textit{E.coli} cells were grown at 37$^{\circ}$C. Fig. \ref{Figure1}\,(left) shows the first and the second calorimetric scan of the membranes in the range from 0 to 90 $^{\circ}$C. The peak attributed to lipid melting is represented with gray shades. One recognizes four peaks above the growth temperature (at 48.7$^{\circ}$, 54.5$^{\circ}$, $\sim$66$^{\circ}$ and $\sim$88$^{\circ}$C), and one peak below growth temperature ($\sim$23$^{\circ}$C). The peaks above 37$^{\circ}$C become smaller in consecutive scans. This suggests that they can be attributed to partially irreversible protein unfolding. The peak at 21.3$^{\circ}$C is  unchanged in the second scan. Its reversible nature suggests that it can be attributed to the lipid melting transition. The transition half width is 14.1$^\circ$ C, which is nearly the same as for BLES and Curosurf. 

Fig. \ref{Figure1}\,(right) shows membranes from \textit{B. subtilis} prepared in the same manner as the \textit{E.coli} membranes. The protein peaks (at 49.7$^\circ$, 60.4$^\circ$ and 69.8$^\circ$C) disappear in the second upscan due to irreversible unfolding. The lipid transition temperature is about 13.6$^\circ$C and the half widths is 17.1 degrees. This peak is still present in the second complete upscan and is reversible. Thus, transitions in \textit{B. subtilis} membranes are similar to those of the previous biological preparations in respect to half width.

Fig. \ref{Figure1} (center) shows the lipid peak of \textit{E.coli} membranes at two different pressures (1 bar and 180 bars). The position of the low temperature peak is shifted by 3.45 degrees towards higher temperature upon application of excess hydrostatic pressure. Using eq. (\ref{lung_1}), one can rescale the temperature axis of the profile measured under pressure such that it is superimposed on the peak recorded at 1 bar. One finds that $\gamma_V=6.5 \cdot 10^{-10}$m$^2$/N, a value that is relatively close to that obtained for DPPC membranes and lung surfactant ($\gamma_V=7.8\cdot 10^{-10}$m$^2$/N) \cite{Ebel2001}. In addition to the reversibility, this strongly suggests that the low temperature peak corresponds to lipid melting. As noted above, the excess volume of most protein unfolding reactions is negative and inconsistent with the observed pressure dependence of the low temperature peak. Overall, we find that the melting behavior of \textit{E.coli} membranes is very similar to that of lung surfactant both in respect to width, transition temperature and to the ratio between excess enthalpy and excess volume. While not showing this here explicitly, we assume that also the total melting enthalpy per gram of the \textit{E.coli} lipids is similar to that of DPPC and lung surfactant, and that one can draw similar conclusions with respect to the temperature dependence of the elastic constants and the time scale of the fluctuations. 
\begin{figure}[b!]
	\centering
	\includegraphics[width=8.5cm]{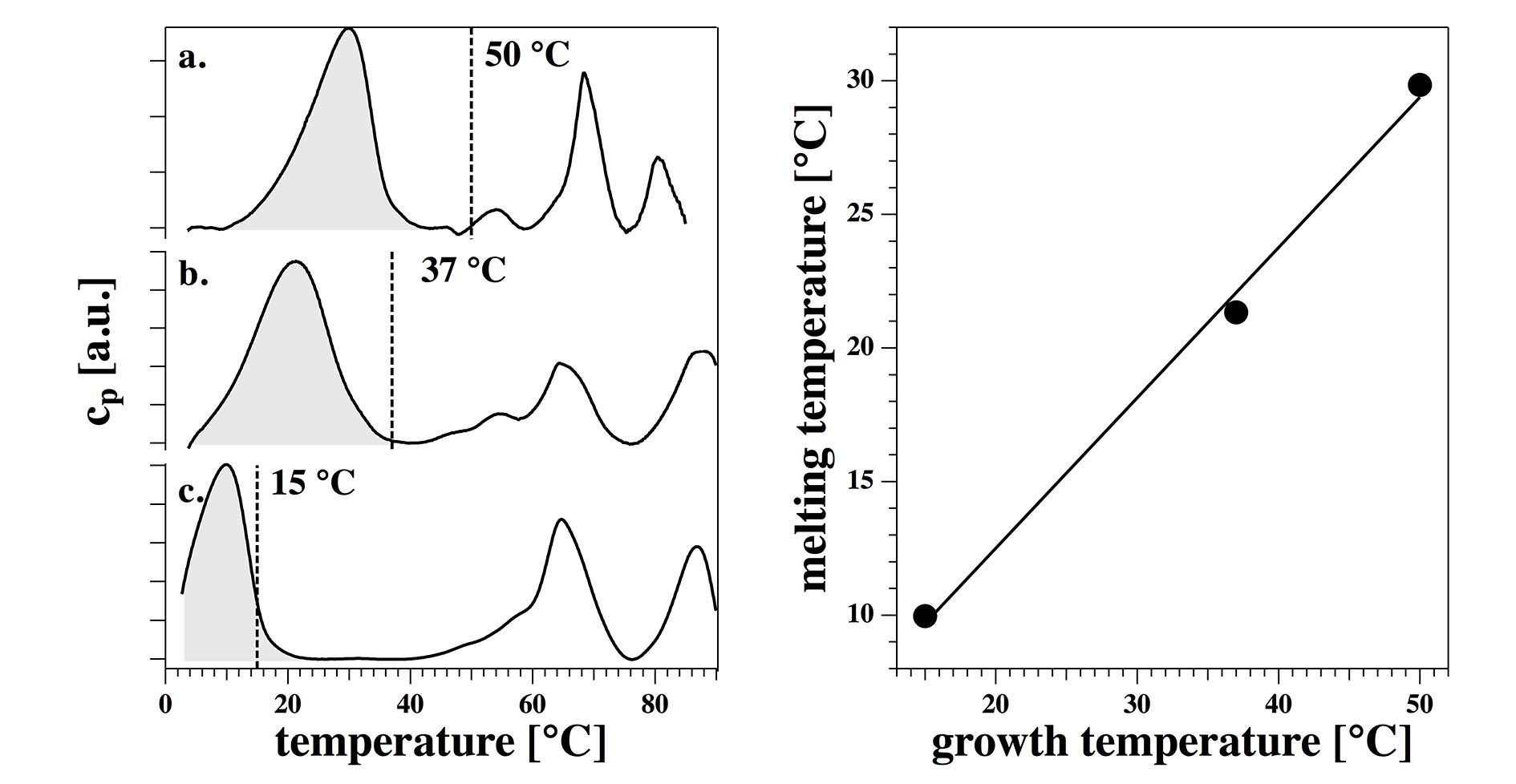}
	\parbox[c]{8cm}{ \caption{\textit{Left: Adaptation of \textit{E.coli} membranes to the growth temperature: 50$^{\circ}$C (a), 37$^{\circ}$C (b) and 15$^{\circ}$C (c). The lipid melting peak is shifted towards lower temperature upon decreasing the growth temperature. The protein unfolding peaks in (b) and (c) are identical while they are slightly different in (a). This indicates that the situations in (b) and (c) correspond to adaptation rather than to mutations. Right: Lipid melting peak maximum as a function of growth temperature.}
	\label{Figure2}}}
\end{figure}
\begin{figure}[b!]
	\centering
	\includegraphics[width=8.5cm]{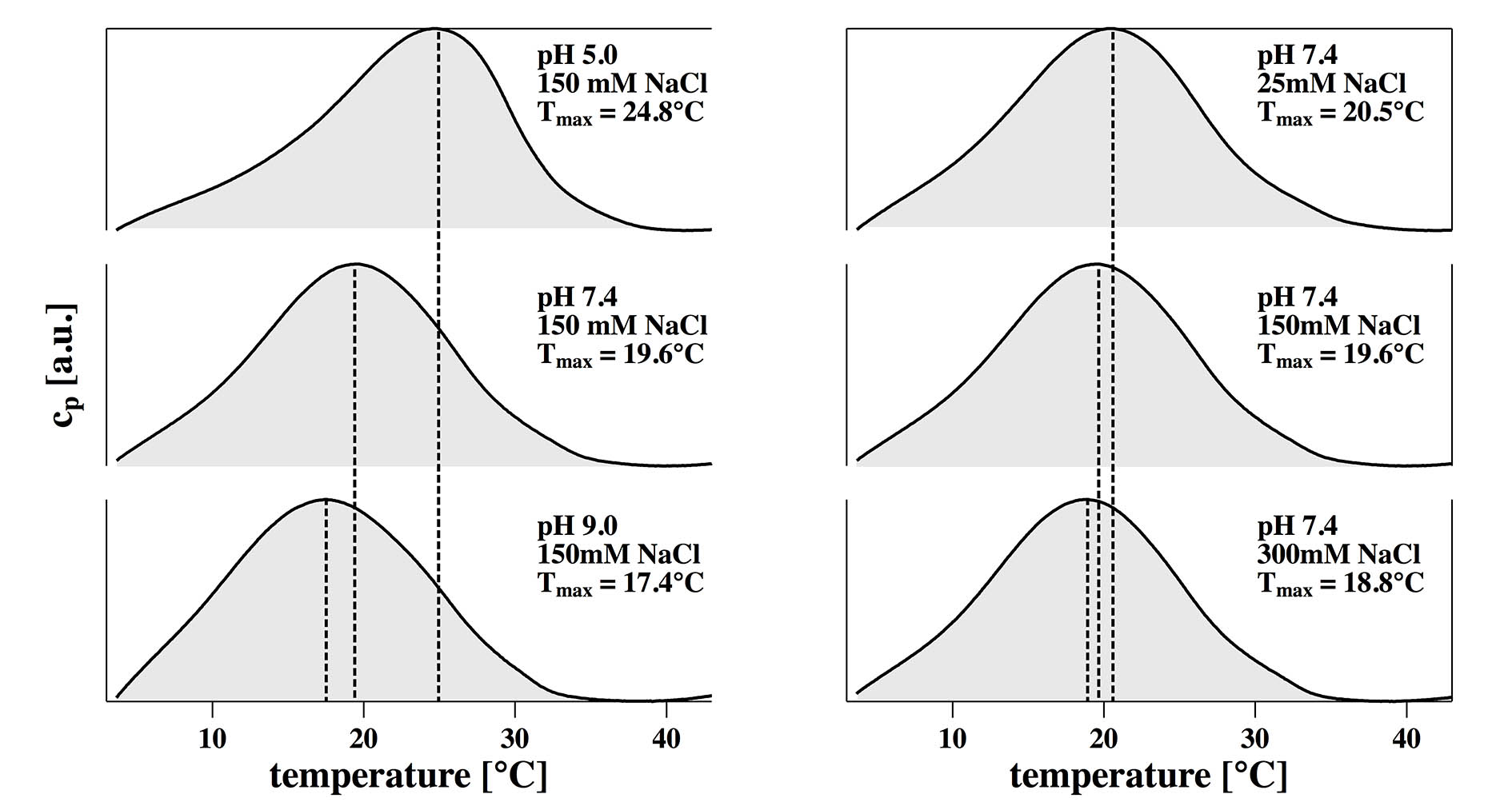}
	\parbox[c]{8cm}{ \caption{\textit{Dependence of melting temperature of \textit{E}.coli membranes on pH (left) and ionic strength (right).}
	\label{Figure1b}}}
\end{figure}
In a further experiment, \textit{E.coli} cells were grown at three different temperatures (15$^{\circ}$, 37$^{\circ}$ and 50$^{\circ}$, respectively, see Fig. \ref{Figure2}). One can see that the melting peak of the lipid membrane (shown in grey shades) moves with the growth temperature. $T_g$. The protein unfolding peaks are found at the same positions for $T_g=15^\circ$C and $T_g=37^\circ$C. This indicates that the cells adapt to different growth temperatures by changing the lipid composition. The protein unfolding peaks for $T_g=50^\circ$C are somewhat different from those at the other two temperatures. This indicates that we may have selected at high growth temperature a temperature-resistant mutant. Evolution of \textit{E.} coli upon exposure to high growth temperatures has in fact been reported previously \cite{Rudolph2010, Guyot2013}. 

Fig. \ref{Figure1b} shows the dependence of the lipid melting transition on pH and NaCl concentration. Conditions are given in the figure. Upon lowering the pH from 9 to 5, the transition temperature increases by 7.6 degrees (Fig. \ref{Figure1b}, left), which indicates that the \textit{E.coli} membrane contains a significant fraction of negatively charged lipids. Fig. \ref{Figure1b} (right) shows the lipid melting peak at three different ionic strength conditions. Increase of the ionic strength from 25 mM NaCl to 300 mM NaCl leads to a slight decrease in transition temperature. \\

\begin{figure*}[htb!]
	\centering
	\includegraphics[width=14.0cm]{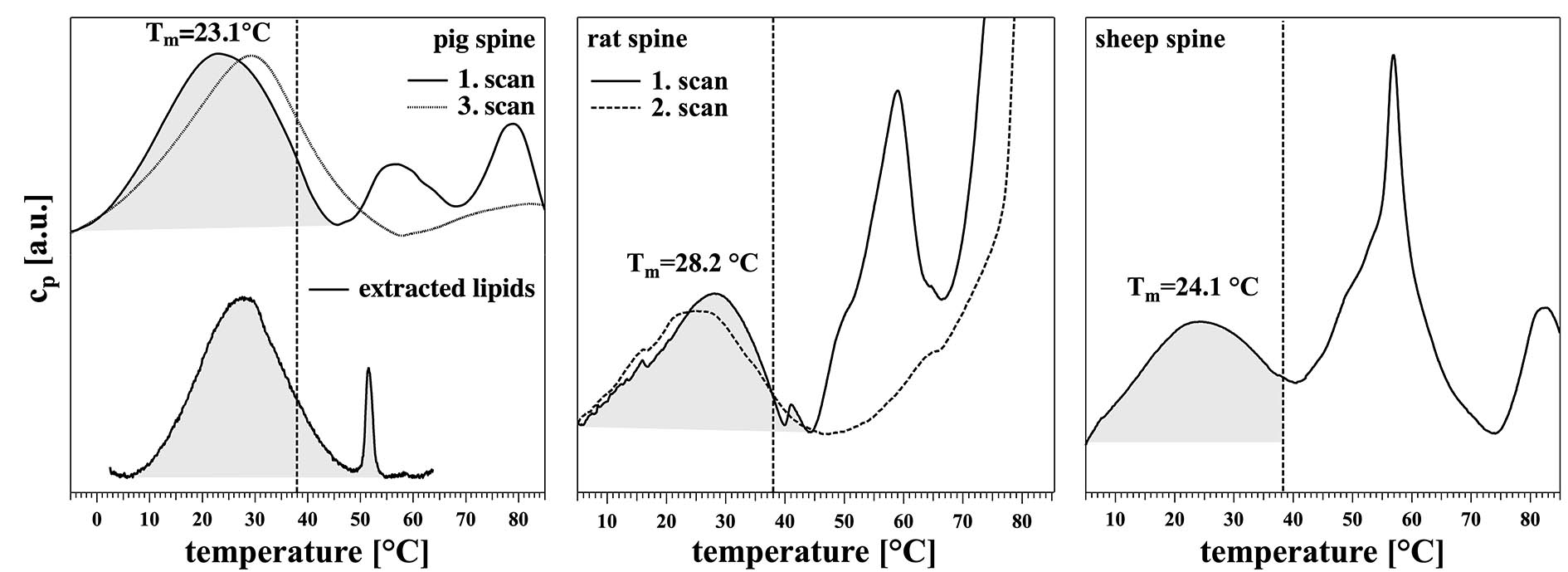}
	\parbox[c]{13cm}{ \caption{\textit{Left, top: Heat capacity profiles of porcine spine membranes. The first scan shows the lipid melting peak and two major protein unfolding peaks. The latter peaks disappear in the third scan due to irreversible denaturation of the membrane proteins. Left, bottom: Heat capacity scan of the extracted membrane lipids of the same membrane preparation. Center: Melting profile of rat central brain membranes. The lipid membrane peak is conserved in the second scan while the protein unfolding is irreversible and the respective protein peaks disappear. From \cite{Madsen2011}. Right: Heat capacity profiles of sheep spine membranes.  From \cite{Tounsi2015, Muzic2016}.}
			\label{Figure_spine_pig_rat_sheep}}}
\end{figure*}
\subsection*{Nerve membranes}\label{nerve}

In the electromechanical theory for the action potential \cite{Heimburg2005c, Heimburg2007b, Heimburg2008, Lautrup2011}, the melting transition plays a central role. Therefore, the investigation of melting profiles of nerve membranes is of particular importance. We have chosen central brain and spinal cord tissues because we assume that here most membranes are related to signal conduction. However, we cannot distinguish different membranes. Our results represent an average over all types of membranes in the tissue. Membranes display a larger density than do soluble proteins. Therefore, our procedure attempts to separate soluble parts from the membranes. The membrane fraction analyzed here contains all membranes from the cells including those of organelles. We assume that a major fraction of the membranes are myelin sheets and axonal membranes. 

Fig. \ref{Figure_spine_pig_rat_sheep} shows the heat capacity profiles for three different membranes preparations from the spinal cord of pig, the central brain (medulla and parts of the spinal cord) from rat and the spinal cord of sheep. The left panel shows the heat capacity profiles of porcine spine membranes. It displays a maximum at 23.1$^\circ$C. There are two further peaks associated with protein unfolding located above physiological temperature that disappear in the second and third complete upscan due to irreversible denaturation. The lipid peak in the third scan seems to be somewhat affected by the denaturation of the proteins, displaying a maximum at about 29$^\circ$C. Proteins aggregating upon denaturation thus diminish lipid-protein interactions. This would fit to the finding that extracted lipids display a c$_p$ maximum at roughly the same temperature (bottom trace of the left panel). In this heat capacity profile one also finds a very sharp low enthalpy peak at around 51.5$^\circ$C. Such a sharp peak is typical for vesicles composed of a pure lipid component. Its origin is not clear and we disregard it due to its low enthalpy content. The center panel of Fig. \ref{Figure_spine_pig_rat_sheep} shows the membrane preparation from rat central brain. Here, we find a lipid membrane melting peak at around 28.2$^\circ$C, and two major protein unfolding peaks at 59$^\circ$C, with a minor shoulder at 50.2$^\circ$C, and 79.9$^\circ$C that disappear in the second complete heating scan due to irreversible denaturation of proteins.  The right hand panel of Fig. \ref{Figure_spine_pig_rat_sheep} shows the calorimetric profile of sheep spine that displays a lipid melting maximum at 24.1$^\circ$C) and protein unfolding peaks at 49.4, 56.7 and 83.3$^\circ$C). The protein peaks also disappear in a consecutive heating scan (data not shown because we could not identify a reasonable baseline). We also investigated small amounts of chicken spine where similar calorimetric events as shown here for the other nerve preparations could be identified (data not shown). Summarizing one can state that the protein peaks are at similar positions in the three preparations, whereas the lipid peaks are not at exactly the same position. This may be an artifact due to heterogeneity of the biological sample preparations.

The half widths of the lipid peaks are of similar order as in the \textit{E. coli}, \textit{B. subtilis} and lung surfactant preparations. Even though we have not been able to measure absolute heat capacities due to the lack of knowledge of the amount of lipids in the sample, we assume that the overall magnitude of the heat capacity and the enthalpy changes are of comparable order as in lung surfactant (see discussion). 


\begin{table}[htb]
	\begin{center}
		\begin{tabular}{llll}
			& pig  & rat & sheep  \\
			\hline
			lipid & 23.1 & 28.2 & 24.1  \\
			protein & - & 50.2 & 49.4  \\
			protein & 56.8 & 59.0 & 56.7 \\
			protein & 78.8 & 79.7 & 82.3 \\
			\hline
		\end{tabular}
		\parbox[c]{8cm}{ \caption{\textit{Major peaks in the heat capacity profiles of native nerve preparations in the first complete heating scan.}}}
		\label{tab1} 
	\end{center}
\end{table}





\section*{Discussion}\label{Discussion}

We showed here that in all biological membrane preparations we have investigated in this study, one finds lipid melting transitions that occur about 10-15 deg below physiological or growth temperature. The preparations were from biological sources as different as Gram-negative and Gram-positive bacteria, lung surfactant and nerve membranes. In a recent publication, we have shown that such transitions also occur in cancer cells of various origins \cite{Hojholt2019}.

The lipid transitions can be identified by various indicators:
\begin{itemize}
	\item In repeated scans over a large temperature interval the protein unfolding peaks are mostly irreversible while lipid melting transitions are reversible.
	\item Lipid transitions display a very different excess pressure dependence as compared to protein unfolding. \linebreak While the excess volume of lipid transitions is positive, leading to an increase of the transition temperature of the order of 1 deg/40 bars (Figs. \ref{Figure_lung_surfactant} and \ref{Figure1}), the excess volume of proteins is usually negative indicating that the temperature of denaturation for proteins is lowered upon increasing pressure \cite{Royer2002, Ravindra2003}.
	\item Comparison to melting profiles of lipid extracts (see Fig. \ref{Figure_spine_pig_rat_sheep}) shows that the melting peaks in the native preparations are similar with respect to transition temperature and transition half width.
\end{itemize}
The enthalpy of the protein unfolding transitions is of similar order as that of lipid melting, although exact ratios depend on the quality of the membrane preparations and the degree to which soluble proteins can be washed out of the samples.

We investigated various membrane preparations: two pre\-parations of lung surfactant, membranes from the Gram-\linebreak negative bacterium \textit{E.coli}, characterized by an inner cytoplasmic membrane and an outer cell membrane, and the Gram-positive bacterium \textit{B. subtilis}, displaying a cytoplasmic lipid membrane and an outer peptidoglycan layer, and three preparations from central brain and spinal cord nerves (pig, sheep and rat). In all of these preparations, we found lipid melting transitions in a range between 10-20 degrees below standard growth or body temperatures. They all share some common features, such as displaying a transition width of about 10-15$^\circ$ and calorimetric events being detectable up to physiological temperature. Thus, the physiological temperature in all of the preparations was found to lie just between the lipid transition and the protein unfolding transitions such that minor perturbations of the membranes will move the membranes into the transition regime. It seems likely that the transition slightly below body temperature is a generic feature of most cell membranes, and this phenomenon may serve important purposes in the functioning of a cell. This feature is conserved from unicellular organisms like bacteria up to the complex nerve system of mammals. 

For two of the biological preparations and the pure lipid DPPC, we measured the pressure dependence of the lipid transition. The pressure dependence yields a coefficient relating the excess volume and the excess enthalpy of the transitions, $\Delta V(T)=\gamma_V \Delta H(T)$ . We have shown previously that the temperature dependence of the excess volume is proportional to that of the excess enthalpy, $\Delta V(T)=\gamma_V \Delta H(T)$ \cite{Ebel2001}. The constant $\gamma_V$ was found to be $7.8\cdot 10^{-10}$ m$^2$/N for DPPC \cite{Heimburg1998, Ebel2001}, $7.6-7.8\cdot 10^{-10}$ m$^2$/N for the lung surfactant preparation (BLES) and $6.5 \cdot 10^{-10}$ m$^2$/N for \textit{E. coli}. A value close to this was confirmed by Pedersen et al in MD-simulations even outside of the transition for DPPC \cite{Pedersen2011}. This implies that the correlation between volume and enthalpy also holds outside of the transition regime. While these values are subject to some experimental deviations for the biological preparations as reflected by the broad transition peaks and uncertainties of baseline subtraction,  they are reasonably close for DPPC, lung surfactant and \textit{E. coli} membranes. Further, we found that the overall excess enthalpy of the transition of lung surfactant was similar to that of DPPC. These two facts have an important consequence: the excess volume compressibility can now be determined from the heat capacity profile when the overall enthalpy of the transition and the magnitude of the excess heat capacity are known: $\Delta k_T^V=(\gamma_V^2 T/V)\cdot \Delta c_p$ \cite{Heimburg1998}. This implies that the membrane volume is more compressible in the melting transition. We assume the same proportionality to be true for the membrane area (excluding the membrane part of proteins): $\Delta A(T)=\gamma_A\cdot \Delta H(T)$, with $\gamma_A = 0.893$ m$^2$/J for DPPC \cite{Heimburg1998}. Thus, we can calculate the changes in area compressibility from the heat capacity to be $\Delta k_T^A=(\gamma_A^2 T/A)\cdot \Delta c_p$ \cite{Heimburg1998}. The latter correlation also allows determination of changes in the bending elasticity  (the inverse bending modulus in Helfrich's theory) to be $\kappa_B=(16/D^2)\cdot \Delta c_p$ \cite{Heimburg1998}. The heat capacity is also correlated to the sound velocity. The velocity of sound in the membrane plane is given by $c=\sqrt{1/\rho_A\cdot \kappa_S^A}$, where $\kappa_S^A$ is the adiabatic area compressibility. It is related to the isothermal compressibility via $\kappa_S^A=\kappa_T^A-(T/V c_p)\cdot(dV/dT_p)^2$. Here, both $\kappa_T^A$, the volume expansion coefficient are simple functions of the heat capacity. Therefore, the sound velocity can be determined from the heat capacity profile. In previous work, we demonstrated that the volume compressibility and sound velocity in lipid dispersion can be calculated from the heat capacity  \cite{Heimburg1998, Schrader2002, Mosgaard2013a}. By analogy, we assume this to be correct in the membrane plane, where the sound velocity in the fluid membrane is given by 176 m/s \cite{Heimburg2005c}. It is significantly slower in the transition regime. 

Our findings have important consequences as summarized below. Many of them have been confirmed for  synthetic lipid membranes. 
\begin{itemize}
	\item at the transition temperature, the volume and the lateral compressibility are at maximum, i.e., the membranes are more compressible within the transition \cite{Evans1974, Heimburg1998}
	\item the bending elasticity is at maximum, i.e., membranes in the transition range are much more flexible than in the fluid or gel membrane \cite{Heimburg1998, Dimova2000}
	\item membrane lifetimes  such as open dwell-time of membrane pores or curvature fluctuation lifetimes are slower in the transition
	\item the sound velocity is in a minimum \cite{Halstenberg1998, Schrader2002}
	\item the magnitude of the effects is similar for all biological preparations shown here because they display similar shapes of their transitions, both with respect to transition temperature and half width of the transition.
	\item since thickness and area of membranes change, transitions can also affect the capacitance of the membrane \cite{Heimburg2012, Mosgaard2015a}. We have previously shown that the capacitance in an artificial membrane can double when going from the gel to the fluid phase \cite{Zecchi2017}.
\end{itemize}
There is good reason to assume that the above effects also exist for biological membranes, but they will be less pronounced because the heat capacity at maximum is lower and the width of the transition is considerably larger.
For the membrane function this implies that
\begin{itemize}
	\item the probability for membrane fusion and fission events is enhanced because it depends on curvature elasticity \cite{Kozlovsky2002}
	\item membrane pores are more abundant because their energy depends on the lateral compressibility \cite{Nagle1978b, Heimburg2010}
	\item lateral sound pulses called solitons can be generated in membranes \cite{Heimburg2005c}
	\item anesthetics, hydrostatic pressure or pH shift the transition and thereby generate changes in compressibility, bending elasticity, pore formation probability \cite{Heimburg2010, Blicher2009, Blicher2013}, open lifetime of membrane pores \cite{Blicher2013} and soliton excitability \cite{Wang2018}. For instance, it has been shown that anesthetics can reduce the open probability of lipid channels in artificial membranes in the complete absence of proteins \cite{Blicher2009, Wodzinska2009}.
\end{itemize}

Our data on \textit{E.coli} membranes show that the lipid melting peak adapts to the growth temperature. If the latter is higher, the lipid transition also moves upwards in temperature in order to maintain a certain distance from physiological temperature. Adaptation of the transition temperature in \textit{E. coli} membranes has already been reported by \cite{Nakayama1980}. It seems likely that the adaptation consists of a change in the fraction of lipids with high and low melting temperatures, or the ratio of saturated to unsaturated chains \cite{Hazel1995}. In particular, for trout livers the lipid composition was shown to be different in winter and summer \cite{Hazel1979}. At 20$^\circ$C, the fraction of saturated lipids is higher, whereas at 5$^\circ$C the fraction of poly-unsaturated lipids is higher. The composition of lipid membranes also adapts to the presence of solvents (among those: acetone, chloroform or benzene) \cite{Ingram1977} or to hydrostatic pressure in deep sea bacteria \cite{DeLong1985}. High pressure increases the melting transition temperature. Consequently, the fraction of unsaturated lipids increases because these lipids display lower melting temperatures which opposes the pressure-effect. For this reason it seems from the few experimental studies that the relative position of the melting transition relative to physiological temperature is a property actively maintained by the organisms.

During a melting transition, membranes display a coexistence of gel and fluid phases that are easy to observe in fluorescence microscopy. This has been well documented for artificial membranes   \cite{Korlach1999, Bagatolli1999, Baumgart2003, Hac2005}, monolayers \cite{Losche1983, McConnell1988, Knobler1990, Gudmand2009} and also for lung surfactant \cite{BernardinodelaSerna2004, BernardinodelaSerna2009}. Phase coexistence is obvious from fluorescence microscopy at the transition temperature, and domains are large. Domain coexistence has frequently been discussed in connection with so-called lipid rafts \cite{Almeida2014}. Rafts are rich in sphingomyelin, cholesterol and certain proteins \cite{Brown1998, Bagnat2000, Bagnat2002, Edidin2003a}. Sphingolipids are known to display high melting temperatures, which are further enhanced by cholesterol. It is therefore tempting to assume that lipid rafts are gel domains swimming in a fluid environment. The fact that physiological temperature is at the upper end of the lipid melting transitions suggests that domains at physiological conditions must be small. This may be the origin of the difficulties to identify rafts at physiological conditions. If rafts are indicators for lipid melting processes, is likely that rafts will be larger and much easier to detect 10-20 degrees below body temperatures. 

We and others have shown in the past that the melting transition of synthetic membranes shifts towards lower temperatures \cite{Johnson1970, Kaminoh1992, Heimburg2007c, Graesboll2014} upon application of general and local anesthetics. This phenomenon follows the simple freezing-point depression law that relates the concentration of anesthetic in the fluid phase to the shift in transition temperature. This simple law has two advantages: 1. It is consistent with the famous Meyer-Overton correlation \cite{Overton1991, Heimburg2007c} and 2. it is drug-unspecific, meaning that it does not depend on the particular chemistry of the anesthetic molecules. In connection with the present investigation we found that the anesthetic pentobarbital lowers transition temperatures in ovine and porcine spine by 2 to 3 degrees when exposed to a buffer with up to 20 mol\% pentobarbital (data not shown). A shift of about 3 degrees was found for 8 mM pentobarbital on DPPC membranes \cite{Graesboll2014}. Thus, the effect reported here is of similar order of magnitude. Since our biological preparations are subject to preparational irreproducibilities and the additional effect of sample aging, we take this as a strong indication that the effect of anesthetics in biological preparations is similar to that in artificial membranes. We will investigate this effect more carefully in future studies. Alltogether, our results are consistent with the finding that alcohols from ethanol to decanol lower the critical temperature of plasma membranes from rat leukemia cells \cite{Gray2013}.

An important consequence of the heat capacity maximum is the accompanying increase in the elastic susceptibilities, leading to more flexible and compressible membranes but also to  decrease in the sound velocity along membrane cylinders. This temperature and density dependence of the lateral sound velocity is the key element of the soliton theory for nerves that reproduces many properties of action potentials \cite{Heimburg2005c, Heimburg2007b, Heimburg2008, Andersen2009, Lautrup2011}. In this model for signal propagation, the nerve pulse is a solitary pulse in which the lateral density and  membrane thickness increases transiently. Both effects have been found in experiments \cite{Tasaki1989, Iwasa1980a, Iwasa1980b, GonzalezPerez2016}. Recently, a theory was put forward that explains how anesthetics can change the stimulation threshold for these density pulses \cite{Wang2018}.


\section*{Conclusions}\label{Conclusions}
We have shown in this work that the lipid melting transitions exist in various biological membranes in vitro. They influence various mechanical features, which as a consequence influence membrane properties such as vesicle fusion probability, pore formation probabilities and their lifetimes, and the generation of solitary pulses in axonal membranes. Since the position of a transition depends on the presence of drugs, pH, pressure, or the presence of proteins, nature gains a powerful tool to influence cell surface properties via the control of macroscopic thermodynamic properties of the biological membrane as a whole.


\vspace{1cm} \noindent\textbf{Author contributions:} 
TM and FT prepared sheep and pig membranes from spinal cord and investigated them calorimetrically, SM prepared and investigated rat membranes from central brain, DP performed some of the calorimetric E.coli and B.subtilis measurements, MK prepared all E.coli and \linebreak B.subtilis membranes. TH performed \textit{E.coli}, \textit{B.subtilis} and lung surfactant experiments and wrote the article. The article was proofread and commented by the other authors.

\vspace{1cm} \noindent\textbf{Acknowledgments:} 
M. Konrad acknowledges continuous \linebreak support by the Max-Planck-Institute for Biophysical Chemistry. We thank Niels Olsen from Rigshospitalet Copenhagen for donations of rat brains. We thank S\o ren Thor Larsen (Haldor Topsoe A/S, Technical University of Denmark and University of Copenhagen) for a donation of Curosurf lung surfactant, and Fred Possmeyer for a donation of lung surfactant preparations (BLES). A few data sets were used in a different context before. Two of the scans in Fig. \ref{Figure2} were used in \cite{Heimburg2007a}. The scans in the right panel of Fig. \ref{Figure_lung_surfactant} were used in \cite{Ebel2001}. We added them for completeness of the discussion. This work was supported by the Villum foundation (VKR 022130).


\small{

}

\newpage
\onecolumn

\newcommand{\beginsupplement}{%
	\setcounter{figure}{0}
	\renewcommand{\thefigure}{S\arabic{figure}}%
}
\section*{Supplement}\label{Supplement}
\subsection*{Baseline removal}

\beginsupplement
\begin{figure*}[h!]
	\centering
	\includegraphics[width=14.0cm]{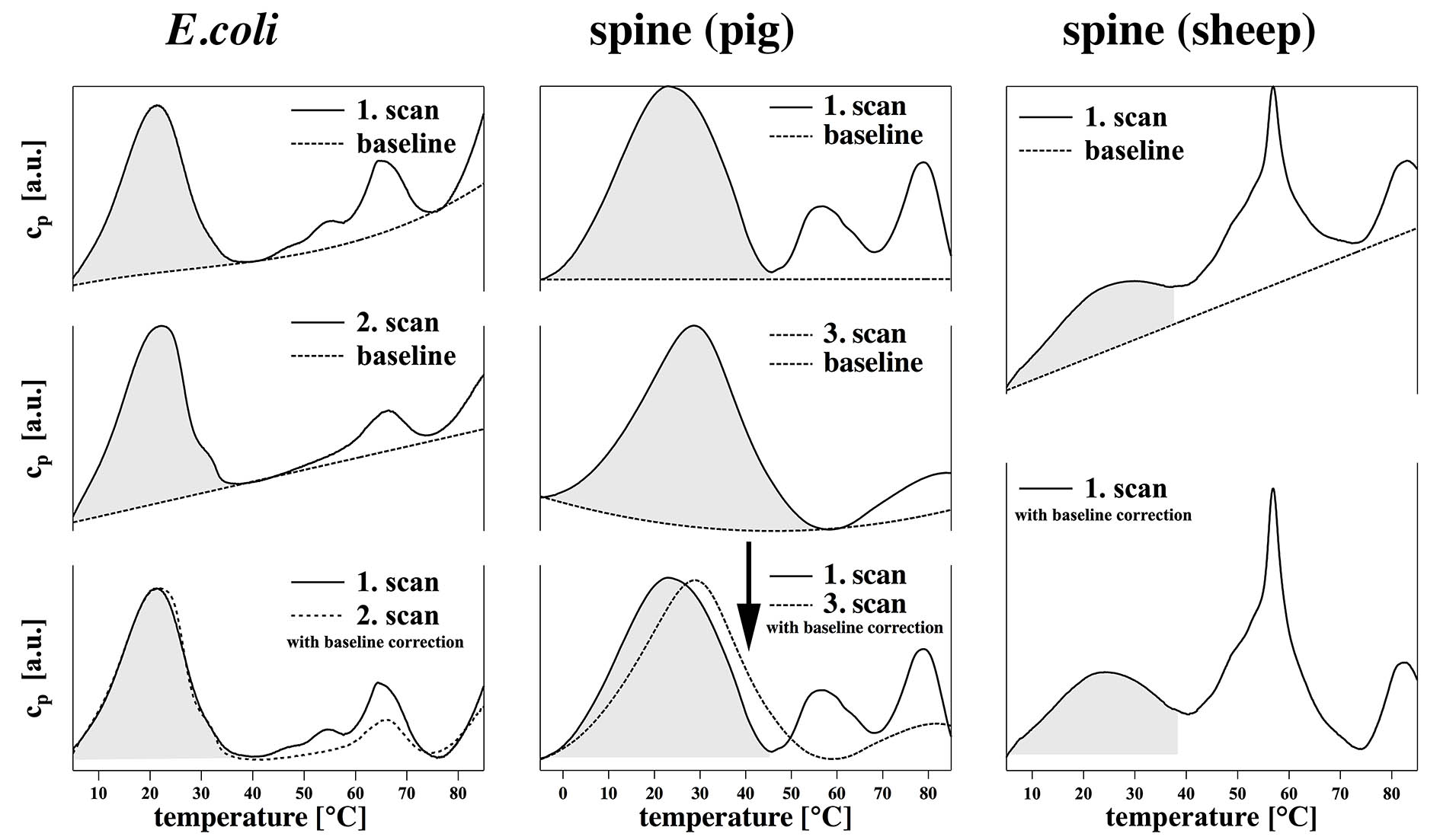}
	\parbox[c]{13cm}{ \caption{\textit{Baseline correction procedure. Left: Scan of E.coli membranes without baseline correction, and subjectively chosen baseline for the first and the second scan, and the baseline corrected scans (bottom). Center: Baseline subtraction for porcine spine membranes. Right: Baseline subtraction for sheep spinal cord membranes.}
			\label{Figure_S1}}}
\end{figure*}
The determination of a calorimetric baseline is difficult for weak signals. It is to a certain degree subjective because calorimetric events span from the lowest temperature to the highest temperature (typically from 0 degrees to 95 degrees centigrade) and there are no obvious reference point in the traces as it is the case for single lipid traces. Especially at the lowest and highest temperature the $c_p$-values are probably subject to high error. It cannot generally be expected that the peaks always begin at the lowest calorimetric temperature and they may extend into the negative temperature regime. It is therefore helpful to use the lowest possible temperature for the start of an up-scan, as done below for the porcine spine membranes, which were recorded in the presence of glycerol.  We have tried to be as conservative as possible with the removal of a baseline. Examples are given in Fig. \ref{Figure_S1}. For the given reasons one should consider the calorimetric traces rather as a qualitative rather than as exact measures.


\end{document}